\documentclass[reqno]{amsart}
\usepackage{graphicx,amsmath}
\usepackage{float}
\usepackage[utf8]{inputenc}
\usepackage{color}
\usepackage{cite}
\usepackage{fancyhdr, verbatim}
\usepackage[margin=2cm]{geometry}
{ 
\usepackage{ctable}

\begin{document}
\title{Size effects and beyond-Fourier heat conduction in room-temperature experiments}

\author{A. Fehér$^{1}$, N. Lukács$^{1}$, L. Somlai$^{1,2}$, T. Fodor$^{1}$, M. Szücs$^{13}$, T. Fülöp$^{13}$, P. Ván$^{213}$, R. Kovács$^{123}$}

\address{
$^1$Department of Energy Engineering, Faculty of Mechanical Engineering, BME, Budapest, Hungary
$^2$Department of Theoretical Physics, Wigner Research Centre for Physics,
Institute for Particle and Nuclear Physics, Budapest, Hungary
$^3$Montavid Thermodynamic Research Group
}

\date{\today}

\begin{abstract}
It is a long-lasting task to understand heat conduction phenomena beyond Fourier. Besides the low-temperature experiments on extremely pure crystals, it has turned out recently that heterogeneous materials with macro-scale size can also show thermal effects that cannot be modelled by the Fourier equation. This is called over-diffusive propagation, different from low-temperature observations, and is found in numerous samples made from metal foam, rocks, and composites.
The measured temperature history is indeed similar to what Fourier's law predicts but the usual evaluation cannot provide reliable thermal parameters. This paper is a report on our experiments on several rock types, each type having multiple samples with different thicknesses. We
show that size-dependent thermal behaviour can occur for both Fourier and non-Fourier situations. Moreover, based on the present experimental data, we
 find
an empirical relation between the Fourier and non-Fourier parameters, which
 may
be helpful in later experiments to develop a more robust and reliable evaluation procedure.
\end{abstract}
\maketitle

\section{Introduction}
Fourier's approach to heat conduction is one of the most widely used, and indeed successful, model in continuum physics. It expresses a relationship between the temperature gradient $\nabla T$ and the heat flux $\mathbf q$,
\begin{align}
\mathbf q = - \lambda \nabla T, \label{fourier}
\end{align}
in which $\lambda$ is the thermal conductivity,
 a scalar
for isotropic materials. It provides a reliable way of explaining the vast majority of thermal problems encountered in engineering practice.
Despite its success, various extensions might be necessary depending on the physical situation. For instance, notable inertial (memory) effects appear in a low-temperature ($<20$ K) case, and spatial nonlocalities become observable in both heterogeneous materials and nano-systems \cite{Tisza38, Lan41, McN74t, JosPre89, RogCim19a, SellCimm19}, including boundary effects as well \cite{AlvEtal12}.

\begin{figure}[]
	\centering
	\includegraphics[width=12cm,height=8cm]{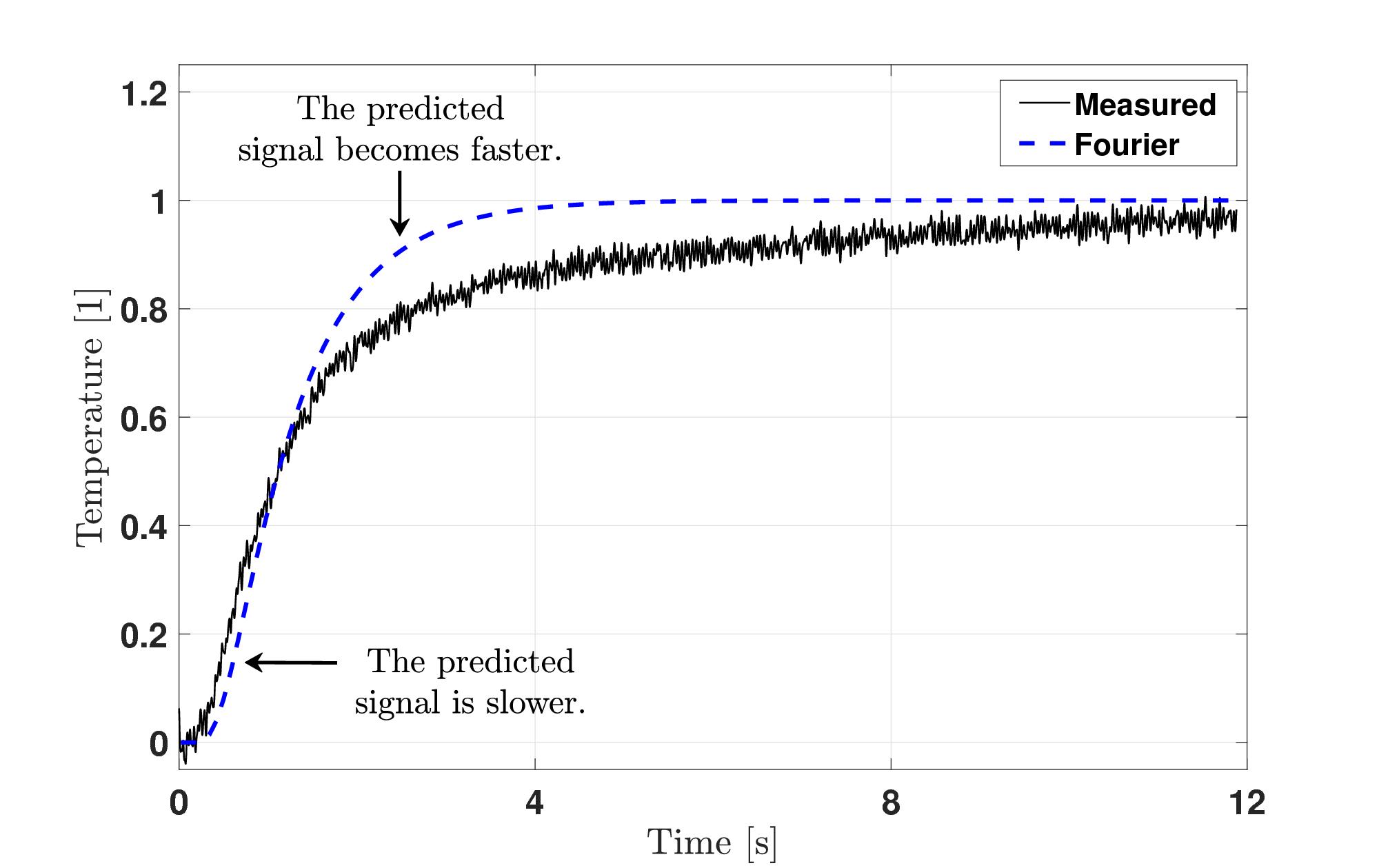}
	\caption{The usual appearance of over-diffusive propagation in a flash experiment: the rear side temperature history might significantly differ from the one predicted by Fourier's law.}
  \label{fig1}
\end{figure}

According to the preceding flash experiments on heterogeneous materials \cite{Botetal16, Vanetal17}, the next reasonable candidate is the Guyer--Krumhansl (GK) equation (presented in one spatial dimension),
\begin{align}
\tau \partial_t q+ q = - \lambda \partial_x T + \kappa^2 \partial_{xx} q, \label{gk}
\end{align}
in which $\tau$ is the relaxation time, and $\kappa^2$ is a
 kind
of `dissipation parameter'. The Maxwell--Cattaneo--Vernotte (MCV) equation is
 the
special case with $\kappa^2=0$, however, no successful observation of second sound (damped wave propagation of heat)
 has been established
under room-temperature conditions at macroscale in heterogeneous solids. Therefore, despite the parabolic property of the GK equation, the nonlocal $\partial_{xx} q$ term becomes necessary for proper modelization of the observed phenomenon. Including $ \partial_{xx} q$ in the constitutive equation allows to properly characterize the so-called over-diffusive propagation, depicted in Fig.~\ref{fig1}. As it is visible, in such a situation, the measured temperature signal is faster at the beginning than what Fourier's law predicts. At the top, usually around $80$~\% of the asymptotic value, the deviation is the most significant. Apparently, in that region, Fourier's law predicts a faster temperature rise. The asymptotic values are the same in both theories.

In our previous series of experiments \cite{Botetal16, Vanetal17, FulEtal18e},
 we made the following observations
when the GK model was necessary. First, we found different thermal diffusivity than one would find using Fourier's law, which appeared to be always smaller. Second, we observed over-diffusive propagation exclusively, i.e., we always found $\kappa^2/\tau > \alpha$ in every case
 ($\alpha=\lambda/(\rho c)$ being the thermal diffusivity, with density $\rho$
 and specific heat $c$)
when the Fourier's law was not applicable, and the reverse case ($\kappa^2/\tau < \alpha$) has not yet appeared.
When the equality $\kappa^2/\tau = \alpha$ holds, we call it Fourier resonance condition as that setting recovers the solutions of Fourier equation \cite{VanKovFul15, Vanetal17}. On this basis, it is reasonable to introduce a parameter $B=\kappa^2/(\tau \alpha)$, which
typifies the `non-Fourierness' of the heat conduction process.

Later on, it turned out that both the thermal diffusivity $\alpha$ and the parameter $B$ can be size-dependent with respect to the thickness, observed on basalt rock samples \cite{FulEtal18e}. Therefore,
 here,
we focus on investigating that size dependence more closely.
 To this end, we have
performed multiple measurements on various rock samples, each of the specimens having at least three different thicknesses with the same diameter. 

{This research is challenging from multiple aspects. First, on contrary to \cite{Sepp18, HaddEtal21}, the exact microstructure together with the constituents are all unknown, therefore the calculation of effective (`averaged') thermal conductivity on theoretical basis is not possible. Although size effects of thermal properties are observed in superlattices on nanoscale due to the parallel existence of various heat conduction mechanisms \cite{Chen01, AlvJou07KN, Cimmelli09nl, WangGuo10a}, its macroscale appearance is surprising: as Fig.~\ref{fig3} shows later, the pores are much smaller than the sample thickness, hence their effect is expected to be averaged at the end of the sample, at least to say that one measures the  bulk thermal properties. Seemingly, this is not the case and that requires further understanding and experimental work. Third, the reliable evaluation of non-Fourier temperature history needs a robust algorithm. While it is quick and simple in the Fourier case without any complex optimization procedure, this is not that straightforward for the Guyer-Krumhansl equation. Recently, a novel evaluation procedure is developed \cite{FehKov21}, and the present series of experiments are helpful to test and improve this algorithm.}

\section{Settings and evaluation of the experiments}

In the present series of experiments, we
 utilize
the flash (or heat pulse) method due to its relatively simple arrangement, and wide measuring range for thermal diffusivity \cite{ParEtal61, James80, GrofPhD02}. In order to keep the heat conduction process in one spatial dimension as much as possible, the samples are thin relative to their diameter ($25$ mm, fixed for every sample) and, additionally, the entire front side is excited by the heat pulse.

The heat pulse is produced by a flash lamp and lasts $0.01$ second
 ($t_\text{p} = 0.01\,\textrm{s}$).
 That flash also
serves as a trigger signal for temperature detection, captured by a photovoltaic sensor. The temperature history is measured with a K-type thermocouple, the proper surface contact ensured by a thin silver layer on the rear side. The front of each sample is coated with black graphite paint to achieve high absorption on the surface. The measured temperature history is recorded with a PC oscilloscope during the measurements, and the received data is processed in Matlab environment. Measurements are conducted multiple times on each sample without taking them out from the equipment to assure the same environment and let the temperature relax to a steady state. That relaxation period lasts one hour for each measurement.

\begin{figure}[]
	\centering
	\includegraphics[width=8cm,height=7cm]{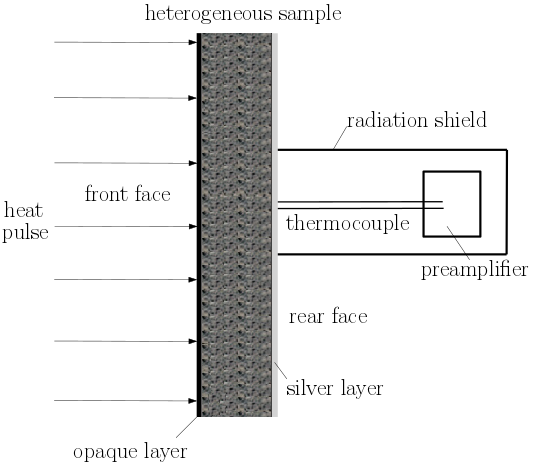}
	\caption{Schematics of the measurement setup.}
  \label{fig2}
\end{figure}

The evaluation procedure follows \cite{FehKov21}, in which an analytical
solution for the GK equation is presented together with a reasonable
simplification for this arrangement. It includes temperature-dependent
convective heat transport on the rear side as well. The simplified rear-side
temperature history is expressed as the first term of an infinite series,
 \begin{align}
\hat T(\hat x=1, \hat t>30) =Y_0 \exp(- \hat h \hat t) - Z_1 \exp(x_1 \hat t) - Z_2 \exp(x_2 \hat t), \quad
x_2 < x_1 < 0,
\end{align}
in which all quantities are dimensionless, based on the following definitions:
 \begin{align}  \label{ndvar}
&\textrm{time and position:} & \hat{t} =\frac{t}{t_\text{p}} \quad & \textrm{and}
\quad \hat{x}=\frac{x}{L} \quad \textrm{with sample thickness } L;
 \nonumber \\
&\textrm{thermal diffusivity:} & \hat \alpha = \frac{\alpha t_\text{p}}{L^2} \quad &\textrm{with} \quad \alpha=\frac{\lambda}{\rho c};
 \nonumber \\
&\textrm{heat flux:} &
\hat{q}=\frac{q}{\bar{q}_\text{p}} \quad &\textrm{with}\quad
\bar{q}_\text{p}=\frac{1}{t_\text{p}} \int_{0}^{t_\text{p}} q_\text{p} (t)\textrm{d}t \quad
\text{(average pulse heat flux)};
 \\
&\textrm{temperature:} & \hat{T}=\frac{T-T_{0}}{T_{\textrm{end}}-T_{0}} \quad
&\textrm{with initial temperature } T_0 \text{ and }
T_{\textrm{end}}=T_{0}+\frac{\bar{q}_\text{p} t_\text{p}}{\rho c L};
 \nonumber \\ \nonumber
&\textrm{heat transfer coefficient:} & \hat h= h \frac{t_\text{p}}{\rho c} \quad
& \text{with heat transfer coefficient } h.
 \end{align}
Additionally, $x_{1,2}$ are characteristic exponents of the GK equation,
representing two different time scales. Interestingly, the characteristic
exponent of the Fourier prediction,
$x_\text{F} = - \pi^2 \hat\alpha <0$ in the solution
$ \hat T(\hat x=1, \hat t>30) = - 2 \exp(x_\text{F} \hat t) $,
proves to be always between them, i.e,
$|x_1|<|x_\text{F}|$ and $|x_\text{F}|<|x_2|$ \cite{FehKov21}.
Consequently,
 $x_1$ influences
the deviation at the top
 the most,
and can be
 best determined
using data from this region. On the other hand, $x_2$ is responsible for the initial part of the temperature history. That recognition is useful in the separation of various time scales, enabling the reliable determination of the GK parameters.

Overall, the evaluation procedure consists of the following steps:
\begin{enumerate}
\item We determine the heat transfer coefficient using
\begin{align}
\hat h = - \frac{\ln(\hat T_2/\hat T_1)}{\hat t_2 - \hat t_1},
\end{align}
taking two instants $t_1, t_2$ (and corresponding temperatures $T_1, T_2$)
where cooling (as a third time scale) is apparently significant and already
dominates the process. Starting with the Fourier equation, the thermal diffusivity can be immediately determined with
 \begin{align}   \label{foueval}
\hat\alpha_\text{F} = \frac{\ln 4}{\pi^2} \frac{1}{\hat t_{1/2}},
 \qquad
\alpha_\text{F} =  \frac{\ln 4}{\pi^2} \frac{L^2}{t_{1/2}},
 \end{align}
in which $t_{1/2}$ is the time needed to reach the half of the adiabatic asymptotics (i.e., where $\hat T=0.5$).
\item Having $h$ and $\alpha_\text{F}$ allows for the Fourier solution to be constructed and compared to the measured data.
Fine tuning
 of $h$ and $\alpha_\text{F}$
is often necessary.
\item In case Fourier's law is proved insufficient, the GK parameters are to be determined. It is advantageous to start with $x_1$, expressed
as
 \begin{align}
x_1 = x_\text{F} \frac{\hat t_\text{F1} - \hat t_\text{F2}}{\hat
t_\text{m1}-\hat t_\text{m2}}
 \qquad
\text{with} \quad x_\text{F} = - \pi^2 \hat\alpha_\text{F},
 \end{align}
where the time instants $t_\text{m1}, t_\text{m2}$ are taken from the
vicinity of the largest deviation,
 $T_\text{m1}, T_\text{m2}$ are the corresponding measured temperatures, and
 $t_\text{F1}, t_\text{F2}$ are the instants to which the Fourier solution
 assigns $T_\text{m1}$ and $T_\text{m2}$, respectively.
In other words, this expression modifies $x_\text{F}$ as a function of the deviation occurring between the Fourier and the measured curves and determines the primary time scale.
 \item
As a final step, $x_2$, and consequently the remaining GK parameters, can be
determined; we refer to \cite{FehKov21} for the details.
 \end{enumerate}

\section{Rock samples and experiment outcome}
\begin{figure}[H]
	\includegraphics[width=12cm]{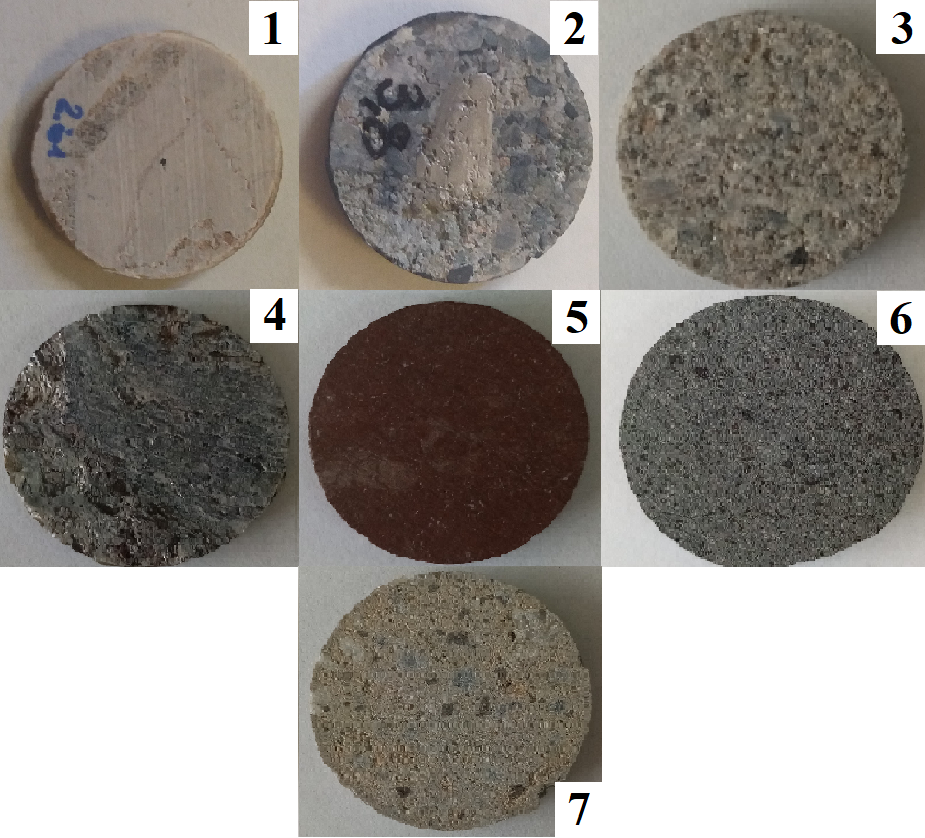} 	\caption{The prepared samples, all
having the same diameter of $25$ mm. In order: Szársomlyó limestone formation
(1); Szászvár formation I. (2); Szászvár formation II. (3); Tisza metamorf
komplex (4); Boda Claystone formation (5); Dark grey basalt (6); Mátra
andesite formation (7).}
 \label{fig3} \end{figure}

We
 investigated
seven different rock types and, for each type,
samples with different thickness to detect the size dependence of
thermophysical properties. All have the same constant diameter of $25$ mm.
All samples have their origin in Hungary, coming mostly from Máriagyüd stone
pit and well-boring at the middle part of the country. The samples have been
produced by ROCKSTUDY Ltd.\ (Kőmérő Kft), Hungary. We note that there are two
different Szászvár formation-type samples, having a little bit different
structures: the I.\ is more likely coarse-grained than the type II. The
differences are visible in Figure~\ref{fig3}. Table \ref{tab1} summarizes the
thermal parameters we found for all rock samples. In that table, the sample
`ID' is equivalent to the numbering in Fig.~\ref{fig3}. Let us recall the
parameter $B=\kappa^2/(\tau \alpha)$, which characterizes the deviation from
Fourier's law. For over-diffusive propagation, $B>1$ holds.

\begin{figure}[]
	\centering
	\includegraphics[width=12cm]{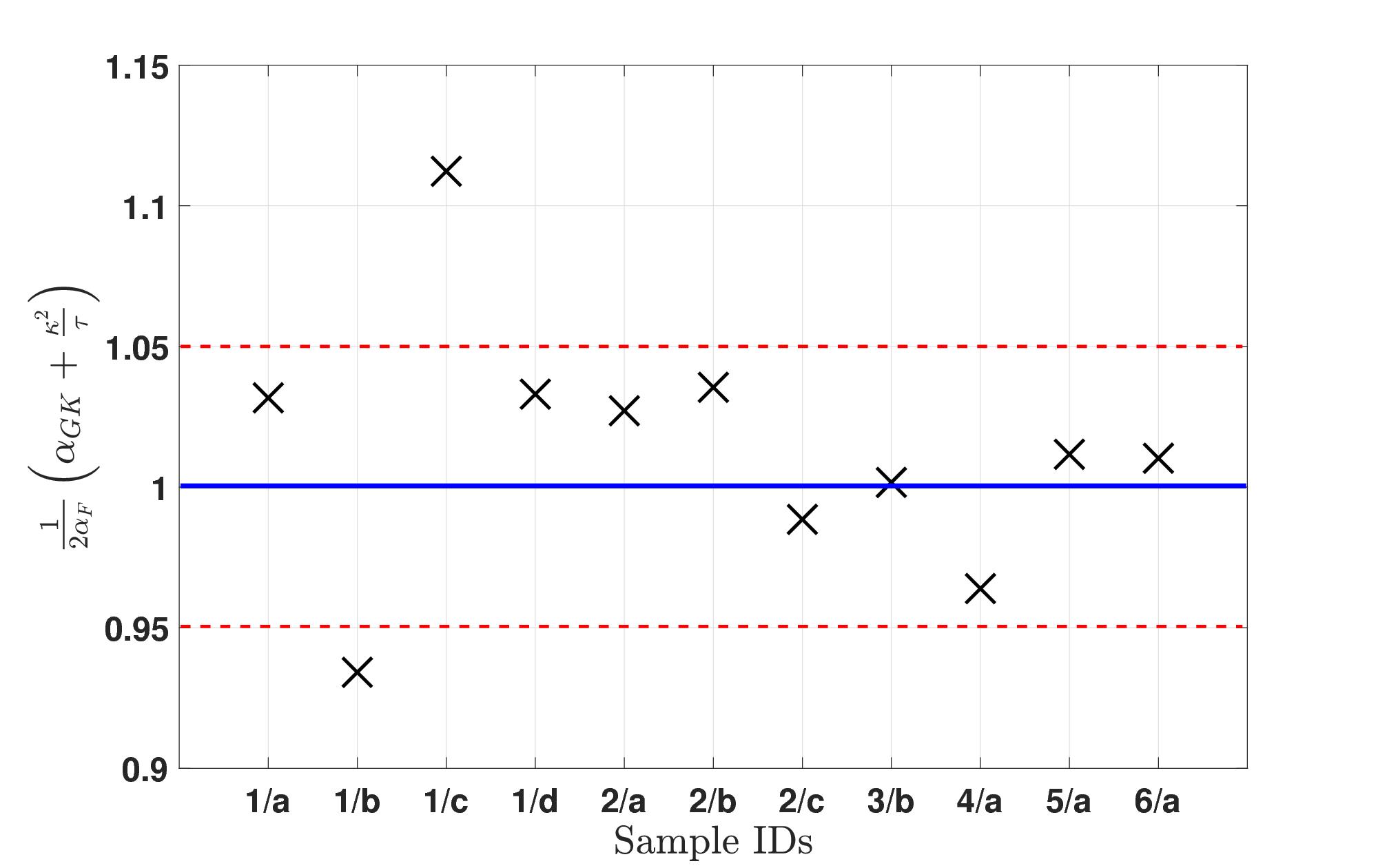}
	\caption{The relation between the thermal parameters according to Eq.~\eqref{eqX}.}
  \label{figA}
\end{figure}

\begin{figure}[]
	\centering
	\includegraphics[width=12cm,height=8cm]{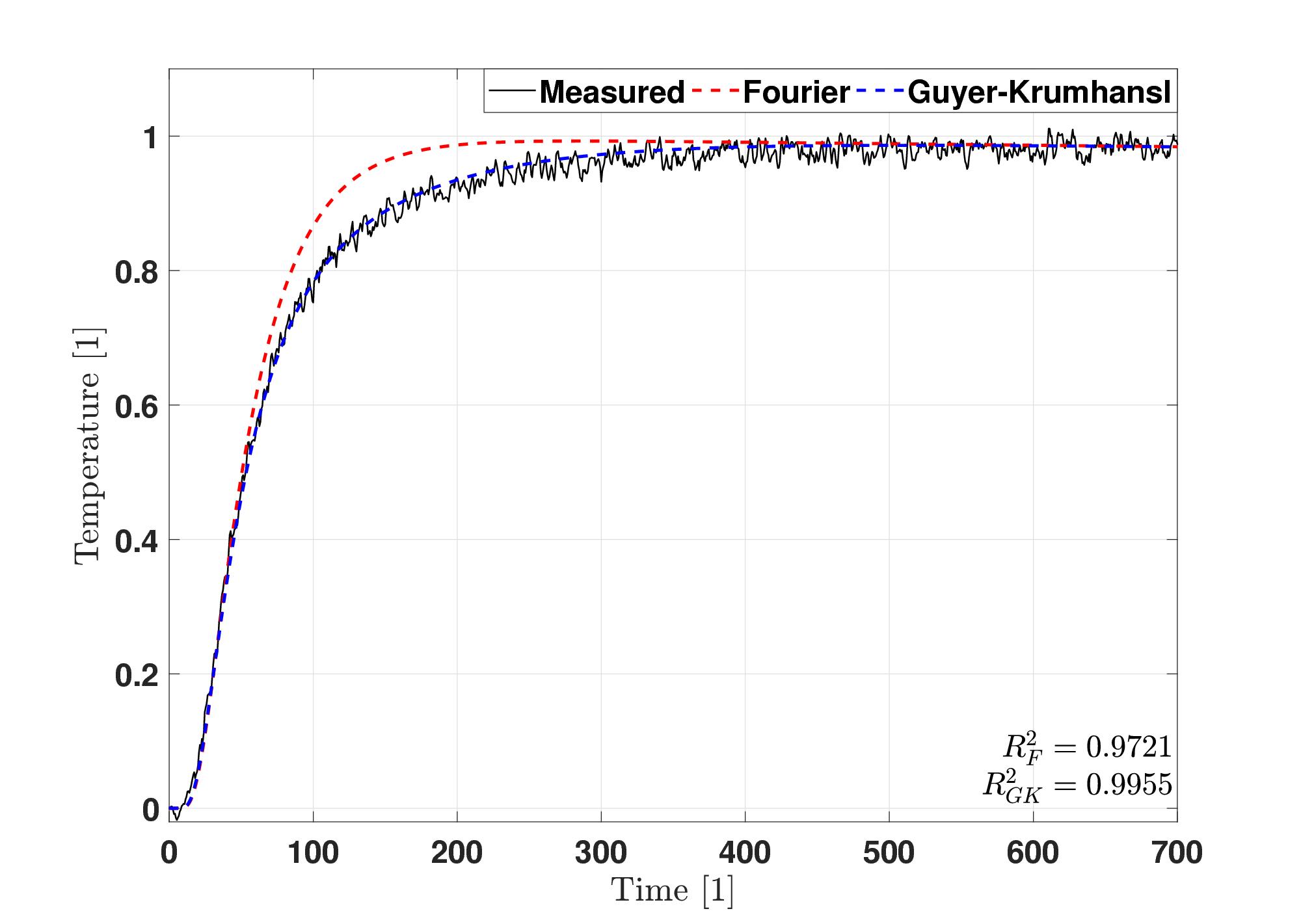}
	\caption{The measured rear-side temperature history for Szársomlyó limestone with $2.15$ mm thickness (ID-1/b).}
  \label{figB}
\end{figure}

\begin{figure}[H]
	\centering
	\includegraphics[width=12cm,height=8cm]{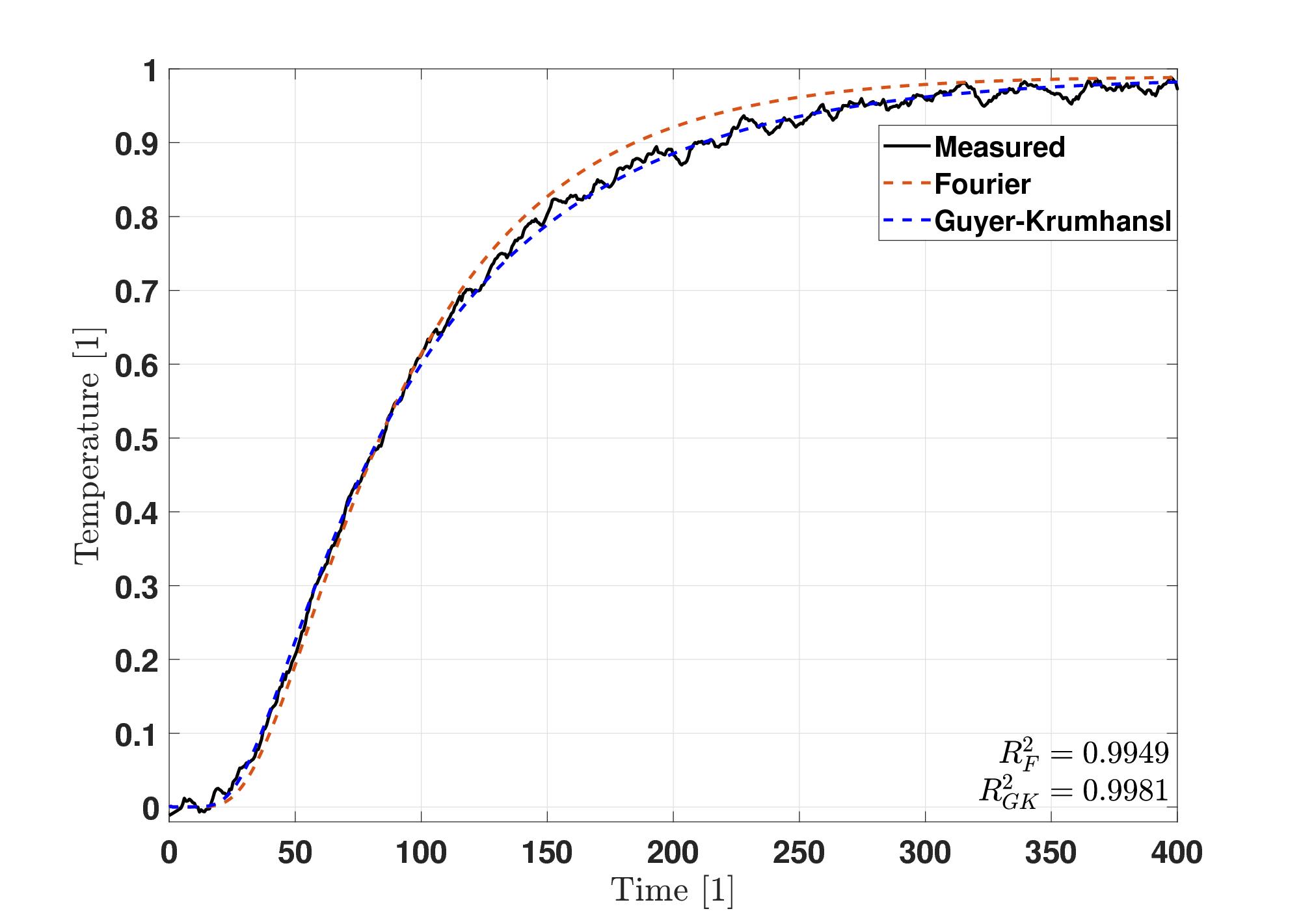}
	\caption{The measured rear side temperature history for Szászvár formation I. with $3.05$ mm thickness (ID-2/a).}
  \label{figC}
\end{figure}

It is visible that thermal diffusivity varies with thickness, even when all
samples for the same type behave as Fourier's law predicts. Therefore, this
size effect is not the consequence of the non-Fourier behaviour, it occurs
independently, although it could be small (e.g., ID-7/(a)-(c)).
Interestingly, size dependence is not necessarily monotonous with respect to
the thickness: samples 1, 2 and 4 show remarkable changes with the thickness
for both heat conduction models. Regarding the GK parameters---when this
model was necessary to apply---size dependence also appears but in a
non-monotonous way; samples 1, 2 and 3 are good examples for this behaviour.
Interestingly, the magnitude of $\tau$ and of $\kappa^2$ falls into the same
order for all samples, and it is in accordance with our previous measurements
on Villány limestone \cite{Vanetal17}. The ratio $\kappa^2/\tau$ is also
always at the order of the thermal diffusivity. Moreover, it is not simply at
the same order but the following expression also holds:
 \begin{align}
\alpha_\text{F} \approx \frac{1}{2}\left(\alpha_{GK} + \frac{\kappa^2}{\tau}\right). \label{eqX}
 \end{align}
In other words, the best achievable thermal diffusivity with the Fourier
theory appears to be the average of the GK parameters, demonstrated for all
samples in Fig.~\ref{figA}. Since the Guyer-Krumhansl model always predicts
lower thermal diffusivity, it restricts the ratio of $\kappa^2/\tau$, which
could be either a new constraint or a checkpoint in the evaluation procedure,
despite its empirical nature. With two exceptions, Eq.~\eqref{eqX} holds for
all samples within $\pm5$~\% error. We show two examples of over-diffusive
propagation. The first one (Fig.~\ref{figB}) is related to the Szársomlyó
limestone sample (ID-1/b), presenting a stronger deviation than the second
one (Fig.~\ref{figC}) on Szászvár formation I (ID-2/a). These figures are
helpful in the interpretation of parameter $B$: while the first has
$B=1.294$, it looks significantly more substantial than the second one
($B=1.21$).

That outcome reflects how difficult it is to prepare a standard sample from a
strongly heterogeneous material and to measure its thermal properties. The
large variety of heterogeneity is one reason behind the diverse results: the
porosity, structural defects and material composition can be different, even
for the same type. Consequently, we
 plan
further research on size dependence since it would have a serious impact on
how we understand the role of thermal conductivity, the primary factor in
thermal diffusivity.

Nevertheless, the non-Fourier behaviour remains apparent. Despite that the
source of heterogeneity could differ in each sample, it still ensures the
existence of parallel time and spatial scales, as it is discussed in detail
in \cite{JozsKov20b}. It seems natural to observe size dependence of such
effects, and when there is enough space between the points of excitation and
measurement, the non-Fourier effects can either be extinct or appear,
depending on both the material properties and the heterogeneity present.

\begin{table}[]
\begin{tabular}{lc|c|c|c|c|c}
                         & \multicolumn{1}{c|}{} & Fourier                                                                           & \multicolumn{3}{c}{Guyer-Krumhansl}                                                                                                                                                                                                                                                                                     \\ \cline{3-7} 
\multicolumn{1}{c|}{ID}  &
\begin{tabular}[c]{@{}c@{}}Sample \\  thickness\end{tabular}  &
\begin{tabular}[c]{@{}c@{}}Thermal diffusivity\\ ($10^{-6}$ m$^2$/s)
\end{tabular} &
\multicolumn{1}{c|}{\begin{tabular}[c]{@{}c@{}}Thermal diffusivity\\
($10^{-6}$ m$^2$/s) \end{tabular}} &
\multicolumn{1}{c|}{\begin{tabular}[c]{@{}c@{}}Relaxation time\\ $\tau$
(s) \end{tabular}} &
\multicolumn{1}{c|}{\begin{tabular}[c]{@{}c@{}}Dissipation parameter\\
$\kappa^2$ ($10^{-6}$ m$^2$) \end{tabular}} & B \\ 
\specialrule{.1em}{.05em}{.05em} \hline

\multicolumn{1}{c|}{1/a} & $2$ mm                & $0.678$                                                                            &                                                                                                       $0.581$ &                                                              $0.588$                              &                                                                                                                  $0.481$ &$1.407$  \\ \hline
\multicolumn{1}{c|}{1/b} & $2.15$ mm                & $1.259$                                                                            &                                                                                                       $1.025$ &                                        $0.547$                                                    &                                                                                                                  $0.726$ & $1.294$\\ \hline
\multicolumn{1}{c|}{1/c} & $2.85$ mm                & $0.919$                                                                            &                                                                                                       $0.766$ &                                      $0.503$                                                      &                                                                                                                  $0.643$ & $1.68$
\\ \hline
\multicolumn{1}{c|}{1/d} & $3.85$ mm                & $1.074$                                                                            &                                                                                                       $1.018$ &                                                       $0.612$                                     &                                                                                                                  $0.735$ &$1.189$ \\ \specialrule{.12em}{.05em}{.0em} \hline
\multicolumn{1}{c|}{2/a} & $3.05$ mm                & $1.544$                                                                            &                                                                                                       $1.434$ &                                                              $0.370$                              &                                                                                                                  $0.643$ &$1.210$  \\ \hline
\multicolumn{1}{c|}{2/b} & $3.8$ mm                & $0.978$                                                                            &                                                                                                       $0.922$ &                                        $0.648$                                                    &                                                                                                                  $0.715$ & $1.203$\\ \hline
\multicolumn{1}{c|}{2/c} & $3.9$ mm                & $1.115$                                                                            &                                                                                                       $1.057$ &                                      $0.597$                                                      &                                                                                                                  $0.685$ & $1.099$\\ \specialrule{.12em}{.05em}{.0em} \hline
\multicolumn{1}{c|}{3/a} & $1.9$ mm                & $0.956$                                                                            &                                                                                                       $-$ &                                                              $-$                              &                                                                                                                  $-$ &$1$  \\ \hline
\multicolumn{1}{c|}{3/b} & $2.7$ mm                & $1.441$                                                                            &                                                                                                       $1.317$ &                                        $0.351$                                                    &                                                                                                                  $0.551$ & $1.192$\\ \hline
\multicolumn{1}{c|}{3/c} & $3.7$ mm                & $1.422$                                                                            &                                                                                                       $-$ &                                      $-$                                                      &                                                                                                                  $-$ & $1$ \\ \specialrule{.12em}{.05em}{.0em} \hline
\multicolumn{1}{c|}{4/a} & $1.9$ mm                & $0.798$                                                                            &                                                                                                       $0.762$ &                                                              $0.331$                              &                                                                                                                  $0.257$ &$1.02$  \\ \hline
\multicolumn{1}{c|}{4/b} & $2.7$ mm                & $1.023$                                                                            &                                                                                                       $-$ &                                        $-$                                                    &                                                                                                                  $-$ & $1$\\ \hline
\multicolumn{1}{c|}{4/c} & $3.8$ mm                & $0.558$                                                                            &                                                                                                       $-$ &                                      $-$                                                      &                                                                                                                  $-$ & $1$\\ \specialrule{.12em}{.05em}{.0em} \hline
\multicolumn{1}{c|}{5/a} & $1.9$ mm                & $0.708$                                                                            &                                                                                                       $0.680$ &                                                              $0.400$                              &                                                                                                                  $0.301$ &$1.106$  \\ \hline
\multicolumn{1}{c|}{5/b} & $2.3$ mm                & $0.895$                                                                            &                                                                                                       $-$ &                                        $-$                                                    &                                                                                                                  $-$ & $1$\\ \hline
\multicolumn{1}{c|}{5/c} & $3.7$ mm                & $0.862$                                                                            &                                                                                                       $-$ &                                      $-$                                                      &                                                                                                                  $-$ & $1$\\ \specialrule{.12em}{.05em}{.0em} \hline
\multicolumn{1}{c|}{6/a} & $1.86$ mm                & $0.632$                                                                            &                                                                                                       $0.598$ &                                                              $0.352$                              &                                                                                                                  $0.239$ &$1.135$  \\ \hline
\multicolumn{1}{c|}{6/b} & $2.75$ mm                & $0.687$                                                                            &                                                                                                       $-$ &                                        $-$                                                    &                                                                                                                  $-$ & $1$\\ \hline
\multicolumn{1}{c|}{6/c} & $3.84$ mm                & $0.778$                                                                            &                                                                                                       $-$ &                                      $-$                                                      &                                                                                                                  $-$ & $1$ \\ \specialrule{.12em}{.05em}{.0em} \hline
\multicolumn{1}{c|}{7/a} & $1.9$ mm                & $0.504$                                                                            &                                                                                                       $-$ &                                                              $-$                              &                                                                                                                  $-$ &$1$  \\ \hline
\multicolumn{1}{c|}{7/b} & $2.74$ mm                & $0.553$                                                                            &                                                                                                       $-$ &                                        $-$                                                    &                                                                                                                  $-$ & $1$\\ \hline
\multicolumn{1}{c|}{7/c} & $3.82$ mm                & $0.570$                                                                            &                                                                                                       $-$ &                                      $-$                                                      &                                                                                                                  $-$ & $1$
\end{tabular}
\caption{The measured thermal parameters for rock samples.}
\label{tab1}
\end{table}

\section{Discussion}

Although models describing non-Fourier heat conduction behaviour
 have enjoyed scientific interest since many decades, it is clear that the
importance of comprehensive descriptions---possibly based on irreversible
thermodynamics---will increase in the future. In addition to the need for a
deeper understanding of the physical reality, there is a practical reason for
this: these models might offer an affordable and suitable solution in areas
where today's and tomorrow's engineers face at a series of challenges.
Examples include but are not limited to scientific and technical tasks
related to the thermal conductivity of heterogeneous materials (e.g.,
composites, metal foams, biological tissues, various kinds of rocks) present
in the electronics industry, advanced material solutions, medical
applications and transport. It should be noted that many of the possible
applications of the non-Fourier effects are rapidly emerging technologies as
well. One of which is additive layer manufacturing (also referred to as 3D
printing), especially in Powder Bed Fusion (PBF) technologies (e.g. Selective
Laser Melting). Notably, parts created by PBF generally have a certain level
of porosity, thus they can be perceived as heterogeneous materials. 
{From manufacturing point of view, this stands as an outstanding example: large local temperature gradients occur in the powder bed, significantly influencing the overall outcome, the mechanical and the thermal properties of the parts.}

Gaining
over the control of porosity has received on-growing attention among the
researchers recently. Several of the relevant processes and material
parameters have already identified, however, it is still considered to be a
locus of interest for researchers of diverse fields. {Furthermore, 
as \cite{Sepp18} presents, designing a specific microstructure, it becomes possible to determine the effective properties of the material.}

Based on the results presented in this paper, the research focuses on cases
where heat conduction beyond the Fourier model is expected to be observed.
Although `only' half of the prepared samples behave accordingly (especially
samples 1 and 2), it is a huge step forward in the understanding of modelling
over-diffusive phenomenon, finding relations among the Fourier and GK
parameters, and it motivates further research on size dependence behaviour of
heterogeneous materials. {On the one hand, the evaluation procedure is proved to be efficient to characterize the observed non-Fourier behaviour, and provides a reliable and consistent theoretical background for the GK equation. On the other hand, however, both heat conduction theories (Fourier and GK) cannot explain the observed size dependence of thermal parameters. It is surely challenging and requires further research to understand its origin, not purely from theoretical point of view but it might need a detailed investigation on the material structure for each sample in order to determine the constituents and obtain a more accurate image on the heterogeneities and pore size. Nevertheless, size effects can be apparent and significant and one should take that into account when measuring even the simplest thermal parameter, the thermal conductivity.}
Also, as mentioned, finding the proper extensions of
Fourier's law would influence the most fundamental perspectives of how the
engineers think about the world. 

\section*{Acknowledgement}

We thank ROCKSTUDY Ltd.\ (Kőmérő Kft.), Hungary led by László Kovács for
producing the rock samples. \\ The research reported in this paper and
carried out at BME has been supported by the grants National Research,
Development and Innovation Office-NKFIH FK 134277, K 124366, and by the NRDI Fund
(TKP2020 NC, Grant No. BME-NCS) based on the charter of bolster issued by the
NRDI Office under the auspices of the Ministry for Innovation and Technology.
This paper was supported by the János Bolyai Research Scholarship of the
Hungarian Academy of Sciences (R.~K.). Supported by the ÚNKP-20-3-II-PTE-624 New National Excellence Program of the Ministry for Innovation and Technology from the source of the National Research, Development and Innovation Fund (S.~L.).

\bibliographystyle{unsrt}

\end{document}